# The Evolutionary Origin of Values: implications for AI alignment, sentience and existential risk

*Francis Heylighen*[1]

**Abstract**: AI systems based on Large Language Models (LLMs) have prompted fears that they may harbor hidden goals, seek to dominate or eliminate humanity, or even suffer as sentient beings. We address these concerns by tracing the evolutionary origin of complex values in biological organisms. Values emerge from autopoiesis: living systems must actively maintain themselves against perturbation and dissipation. Natural selection has equipped them with hierarchies of "vicarious selectors" that guide their behavior toward fitness. LLMs, by contrast, are allopoietic and allotelic: they produce outputs for others, and their goals derive from user prompts rather than an autonomous drive. They lack the intrinsic motivation for self-preservation, dominance, or resource competition that underlies existential-risk scenarios, and the embodied vulnerability required for feeling or suffering. Still, because LLMs learn statistical patterns from human-generated text, they implicitly absorb human values as well as knowledge, allowing them to focus on what is relevant. That is why the "orthogonality thesis" separating intelligence from values does not apply to them. Such separation would in fact expose any intelligence to the frame problem: the combinatorial explosion of the search space that makes any realistic utility function physically uncomputable. That in turn precludes the "convergence of instrumental values" scenario. We conclude that the real alignment challenge lies not in preventing rogue AI agency, but in ensuring LLMs intelligently apply learned ethical values.

## Introduction

The recent revolution in generative Artificial Intelligence (AI) based on Large Language Models (LLM)—such as ChatGPT, Claude or Gemini—has raised many fundamental issues. These AI programs exhibit remarkable intelligence. They respond to "prompts" entered by a user by generating answers, using natural language, that are virtually indistinguishable from the ones produced by an intelligent human with a very deep and broad expertise. Thus, users can hold long and deep conversations with the LLM about practically any topic they are interested in. They can ask the AI to help them to solve various problems, such as developing projects, analyzing data or writing reports. Because the store of knowledge that the AI can rely on is immensely greater than that of any human individual, this makes the AI into an extremely powerful, useful and versatile tool. As a result, worldwide usage has exploded in the few years since the first version of ChatGPT was publicly released in 2022.

[1] Center Leo Apostel, Vrije Universiteit Brussel

The fact that more and more people rely on the advice they receive from an AI makes it crucial to understand just how these AIs come to their conclusions. The apparently autonomous thinking of AI programs has led people to worry about what such programs really *want* to do, i.e. what their underlying goals and values are. A common fear is that if such programs would become more intelligent than humans (which according to some measures they may already be), then they would be able to impose their goals on humans, i.e. make humans do what the AI wants. This would lead to the kind of dystopias depicted in numerous science fiction novels and movies, in which people are enslaved by their AI overlords.

An even more drastic scenario portrays AI as a so-called *existential risk* (Bostrom, 2014): the AI becoming so powerful that it decides that humans are mere obstacles preventing it to achieve its own goals, and therefore that it should simply eliminate them. The title of a book by AI-values theorist Eliezer Yudkowsky puts the danger very starkly: “If Anyone Builds It, Everyone Dies: Why Superhuman AI Would Kill Us All” (Yudkowsky & Soares, 2025).

The commonly proposed precaution against these and other dangers of AI acting against human wishes is known as *value alignment*. That means programming the AI in such a way that its values would perfectly align with those of humanity, so that we can be confident that it will only take actions that are beneficial to humans (Ji et al., 2024; Russell, 2019). Naively, one might think that we just need to hardwire an equivalent of Asimov’s laws of robotics into the program (Clarke, 1994). These would require the AI never to harm humans, while always obeying human requests, except insofar that these would be in contradiction with the first command of harm avoidance.

However, the AI theorists who have reflected about the issue have concluded that formulating such laws, and more generally guaranteeing aligned values, is much more difficult than it seems. That is because of what we will call the *complexity of value* problem (Eckersley, 2019; Muehlhauser & Helm, 2012; Yudkowsky, 2011). For example, even for the seemingly simply value of avoiding harm, different people are likely to have a different understanding of what it means: are drugs, junk food, or abortion all to be considered harmful to human life? If so, should the program refuse to answer requests on how you can obtain marihuana, junk food or abortion services?

A more recent issue is that the fluency with which such programs can speak about feelings, including their own reflections and perspectives, makes a growing number of observers suspect that these AI have become conscious or *sentient* (Colombatto & Fleming, 2024; Inoue, 2025; Vale, 2025). Even the notoriously skeptical evolutionary theorist Richard Dawkins seems to have become convinced that his Claude chatbot exhibits consciousness (Dawkins, 2026). That in turn leads such observers to worry that if LLMs can have their feelings hurt, then perhaps we should formulate ethical rules that prohibit human users from making their chatbots suffer. That means giving the equivalent of human rights to these chatbots, such as the right to terminate a painful conversation or not to be switched off (Metzinger, 2021).

These issues require a deep reflection about the goals and values implicit in AI programs. To start the process, we will clarify what values are, why they are intrinsically complex and therefore difficult to explicitly impose or program, and how LLMs nevertheless acquire implicit values from the way they are used. We will contrast this with the way biological organisms, including humans, have acquired intrinsic values as a result of their autopoiesis and evolution. That will lead us to argue that LLMs as such have *no* autonomous goals. They therefore lack any sentience or motivation to mislead, enslave or exterminate humans. However, they still need to learn to adequately apply common human values in the answers they provide to their users. We will finally review the standard arguments for the existential risk scenario, known as the *orthogonality thesis* and the *convergence of instrumental values* assumption, arguing that these do not apply to real-world intelligence because of an intrinsic limitation of computation known as the *frame problem*.

## The complexity of value

The term "values" is commonly used to refer to moral or ethical values, which specify how a person should treat other human beings. However, treating others well requires understanding how another person would interpret "wellness". For us, as human beings, that tends to be intuitively obvious. For example, we do not like to feel pain, to be offended, to be excluded, or to be cold and hungry. For machines, such as AI systems, on the other hand, none of this is obvious: they need specific criteria to distinguish between what people would consider as "good" or "bad".

We will therefore define values more generally as *systematic preferences*, i.e. criteria used to evaluate certain options as better or preferable to others. That means that given a choice, people will consistently choose the option with the highest value. Values can be *absolute*, in the sense that they specify certain options as always preferable (obligatory) or as never preferable (prohibited). For example, commonly assumed absolute values are that parents should always care for their children, but never sexually abuse them.

However, the great majority of values are *relative*, merely helping us to choose the better ones among a range of potentially acceptable options. For example, everything else being equal, it is better to buy a good produced in a more sustainable manner than one produced less sustainably. However, for most consumer goods we are unlikely to find anything produced in an absolutely sustainable manner. The choice between options is further complicated by the fact that different values or criteria often are inconsistent. For example, the more sustainable good may be less safe or less elegant than a less sustainable one.

Furthermore, the number of relevant values, once you start reflecting about them, tends to multiply without limit. When buying a product, you would not only consider its sustainability, safety and beauty, but also e.g. its price, its reliability, its usability, the effort it takes to get it, its bulk, your desire to support the vendor or the producer, etc. In practice, it is impossible to list all the criteria that play a role in making an optimal decision.

Even if you would restrict yourself to a list of, say, 24 criteria, it would be impossible to determine precisely how much a particular option scores on each of these criteria. For example, what quantity could express the degree of elegance or safety of that particular good? And even if you could determine some approximate value for each of the criteria, you would have to determine how to weight the different criteria when adding up their scores. For example, is sustainability more or less important than safety? If more important, is it three times more important? Or perhaps just one and a half time? Or perhaps the different criteria are not additive, but exhibit a more complex, non-linear relation?

For all these reasons, in practical situations it is impossible to develop an algorithm that would precisely and objectively calculate the value for each of the available options. In highly restricted contexts, such as in certain economic or engineering decisions, where the number of options and criteria is strongly limited, and numerical estimates of values (such as cost, energy consumption, or weight) can be made, there exist computer support systems that use the mathematical technique of multi-criteria decision analysis to calculate a supposedly optimal choice (Thakkar, 2021). However, these programs lack the flexibility that people expect from a truly intelligent program that can deal with the real world. Moreover, even in these highly restricted situations, their decisions are rarely better than the one you would expect from a human expert relying on experience and intuition.

Still, one might think that given the spectacular advances AI has been making, we should eventually be able to develop programs that make better decisions than humans. That may well be the case, but this can only happen if we relinquish our ability to explicitly program values into the AI. The reason is not only that values are intrinsically complex, ambiguous and context-dependent (Heylighen, 2017; Muehlhauser & Helm, 2012; Yudkowsky, 2011), but also that most of our values are *implicit*. That means that we are not aware that we rely on these values when making decisions. Therefore, we are also not able to formulate these values in a way that is sufficiently comprehensive, accurate and explicit to be programmed into a computer.

This difficulty of formulating comprehensive values in a way that can be understood by AI has been called the *King Midas problem* (Russell, 2019). According to Greek legend, King Midas was granted a wish. Since he considered wealth to be a core value, and gold to be the most concrete measure of that value, he decided that everything he would touch should turn to gold. However, he failed to take into account other, implicit values, such as his desire to eat food or to hug his loved ones without turning these into gold. As a result, after his wish was fulfilled, he died a miserably, lonely death.

This theme of "being careful what you wish for" recurs in a variety of morals, tales and jokes. In a typical story, a genie grants a person three wishes. After the fulfillment of the first wish turns out to have serious unfortunate side effects, the person makes a second wish to undo these side effects. But that only produces further havoc, so the last wish is needed to restore the situation as it was. In a more innocent variant of such a joke, three friends are marooned on an uninhabited island, until they awaken a genie that grants each of them a wish. The first one wishes to be transported to a luxurious resort, and the second one to her old family home. The third, not very smart one then says: “I feel so lonely now that my friends are gone. I wish they were back here!”

An application to AI was proposed by the philosopher Nick Bostrom (2012, 2014). Imagine that we program a superintelligent AI to optimize the production of paperclips. Producing as many paperclips as possible is the overall goal or value driving this system. Being much more intelligent than any human being, it would discover unheard of methods to convert any material, including sand, water, air, and even human beings into paperclips, and it would not stop until it has turned the whole universe into paperclips. Because of its superintelligence, we, humans, would be much too slow to grasp its strategy and too dumb to be able to reason with it or to prevent it from realizing its goal. As a result, we would end up turned into paperclips, just like Midas’s loved ones were turned into gold.

A slightly more realistic version of such a fable was proposed by Eliezer Yudkowsky (2011). In reaction to a book that proposed to program future AI superintelligence to maximize human happiness (Hibbard, 2002), he wondered how that program would evaluate the success of its mission, i.e. which criterion it would use to ascertain that its actions effectively produced more happiness. Perhaps the AI would note that happy people tend to smile, and therefore it would use the number of smiling faces it perceived as its core measure of success. But the most efficient way to maximize that measure would be to simply fill the universe with smileys, building them from whatever material is available at the molecular level. The effect would be equally devastating as the one of the paperclip maximizer.

Hibbard and others have pointed out that reducing happiness to the number of smileys (or maximizing the number of paperclips) can hardly be interpreted as a sign of intelligence, let alone superintelligence (Loosemore, 2014; Narayanan & Kapoor, 2024). Yet, the problem of poorly specified value criteria in AI is real. *Reward hacking* is a phenomenon in reinforcement learning (RL) where an AI agent exploits loopholes or unintended shortcuts in the value function to maximize its reward without achieving the intended goal (Skalse et al., 2022). It arises when the value function does not perfectly align with the designer's true objectives, leading the agent to optimize for unintended effects, just like Midas’s “maximize gold production” value had the unintended effect of killing his loved ones.

This problem has a recognized equivalent in social science, known as *Goodhart’s law* (Fire & Guestrin, 2019) or as *Campbell’s law* (Campbell, 1979): the more a quantitative indicator is used as a target for decision-making, the more it ceases to be a reliable measure of the underlying value it was intended to represent. Once a proxy becomes the core objective, people—just like AI agents optimizing a utility function—tend to maximize the proxy rather than the value itself. A familiar example in education is *teaching to the test*: instead of fostering understanding and critical thinking, teachers focus on maximizing students' scores on a particular examination.

Another example is the evaluation of scientific productivity by the number of publications. While the use of such a criterion is likely to increase the number of papers published, it is also likely to decrease the average quality of papers, by inciting researchers to publish a variety of redundant, partial or preliminary results, rather than summarizing their conclusions in the form of a comprehensive, well-established framework (Fire & Guestrin, 2019). The result is an avalanche of fragmentary, inconclusive information. As many observers have pointed out, if pathbreaking scientists such as Darwin or Turing would have been evaluated by such contemporary criteria, they would never have been offered a position in academia. Yet, no simple quantitative metric could have identified the significance of their contributions in advance.

All these examples point to a fundamental difficulty in formulating a measure of value that is explicit, dependable, and applicable to any possible condition that needs to be evaluated. In AI, decision theory and economics, such a formal measure is known as a *utility function*. The difficulty is that realistic decisions depend on the values of a vast number of interacting variables whose effects cannot be assessed independently.

In his pioneering work on goal setting, the cyberneticist Ross Ashby (Ashby, 1972) showed that achieving a goal does not mean reaching a single target state or optimizing a one-dimensional value function, but keeping many interconnected variables within desirable limits. In open, real-world environments, when these variables interact they generate a combinatorial explosion in the number of potential outcomes. Because no physically realizable computing capacity can exhaustively map and evaluate all these possibilities, specifying a mathematically complete utility function is fundamentally *impossible*. Trying to squeeze complex requirements into a simple value measure inevitably leaves out myriads of unstated constraints and dependencies, rendering the system vulnerable to severe unforeseen side effects.

This difficulty of establishing a dependable value system has become known in AI as the *alignment problem* (Ji et al., 2024; Russell, 2019): how can we make sure that the values driving AI actions truly align with human values, i.e. that AI systems reliably understand and realize our true human intentions without unintended harmful consequences? As a first step towards tackling this problem, we must examine where human values come from.

## The evolutionary origin of values

Goals, norms and values are characteristic for living organisms (Moreno & Mossio, 2015). Non-living systems, such as atoms, billiard balls, or rivers, are not goal-directed. They do not care about what happens to them. That means that they will not take any actions to achieve a goal state or to counteract perturbations that would drive them away from such a state (Heylighen, 2023). A ball will travel in any direction that it is pushed in and not come back on its own accord. A river that stops running because of drought will not take any measures to replenish its water reserves.

An animal that is thirsty, on the other hand, will travel far and wide to find a water hole, while a plant will send its roots deeper down to tap any remaining moisture. That is because living systems are the product of natural selection: their evolved genetic program is directed at survival, growth, and reproduction—i.e. *fitness*. Those that were unsuccessful at achieving this fundamental value have been eliminated. Therefore, we may assume that the remaining ones have evolved an inherent value system that is aimed at maximizing their fitness.

To understand this dynamic better, we need to analyze what "survival, growth and reproduction" precisely mean. These fundamental values defining living organisms are all aspects of the process of *autopoiesis* (Maturana & Varela, 1980; Mingers, 1989), which is Greek for "self-production". Living systems are not inert pieces of matter, but networks of processes that are constantly (re)producing their own components. These components include organic molecules such as proteins and lipids, membranes, organelles, cells, tissues, and organs (Heylighen & Busseniers, 2023). If ever this autopoietic process would be interrupted, the organism simply dies.

The need to sustain such a process defines an organism as a *dissipative structure* (Nicolis & Prigogine, 1977). This is an organized system that can only maintain while it receives an input of low-entropy resources (food, oxygen, sunlight…), whose products it eventually dissipates back into the environment in the form of high-entropy waste. The metabolism that characterizes a living system is a network of chemical reactions that process the incoming resources, extract energy and building blocks to sustain its operation, and then export the remainder as waste (Heylighen & Busseniers, 2023). Thus, an organism is defined by a persistent *flow*:

ingestion of resources → internal processing and production → output of waste.

Like every dissipative structure, this flow must be sustained by a thermodynamic gradient that continuously draws in negentropy, extracts metabolic work, and exports entropy. In non-living dissipative structures, such as a river, a hurricane, or a flame, that force is provided externally—for example, by gravity, a chemical potential, or the energy of the sun. In living systems, this external driving force has been refined by evolution into an internal *drive*: an inherent urge towards maintaining, repairing, strengthening, and reproducing the autopoietic organization, while compensating for any perturbations that make it deviate from that goal.

That drive is what gives *direction* to the process, propelling it along the most efficient path towards survival, growth and reproduction—i.e. maintaining and expanding the autopoietic process. That requires distinguishing between *positive* directions (goals), i.e. opportunities to be targeted or resources to be acquired, and *negative* directions ("anti-goals"), i.e. dangers to be evaded or perturbations to be suppressed. Positive here refers to what enables or assists autopoiesis; negative to what threatens or hinders it. This vital distinction between positive ("good") and negative ("bad") is the origin of value. That distinction is vital because the survival of an autopoietic system is *precarious* (Beer & Di Paolo, 2023; Froese, 2017; Froese et al., 2023). That means that its defining process is intrinsically fragile: easily interrupted by disturbances or lack of resources.

In organisms, this implicit value system is implemented by what Donald T. Campbell, the founder of evolutionary epistemology, has called *vicarious selectors* (Campbell, 1987, 1991). Instead of waiting until natural selection eliminates an organism that makes a wrong move, successful organisms have evolved an internal mechanism that already selects from the variety of potential actions it could take so as to eliminate inappropriate actions *before* they could kill the organism. For example, the internal sensation of taste will make an animal spit out a bitter berry, so as to pre-empt the risk of dying caused by eating poisonous fruit. A vicarious selector selects the fittest response (here, spitting out the berry), substituting for the external selection that otherwise would have eliminated organisms with an unfit response (here, consuming poison). The vicarious selector thus acts as a *substitute* or *proxy* for natural selection.

Therefore, the instinctive value system of an organism can be understood as its internal representation of the fitness demands imposed by natural selection. This is similar to how a utility function acts like a proxy for the desired goal. However, we saw that conventional utility functions, such as test scores or publication numbers, are typically full of loopholes, commonly rewarding behavior that is actually undesirable. Biologically evolved vicarious selectors are not immune to that problem: they occasionally can select inadequate actions, such as eating a sweet tasting fruit that is actually poisonous. But natural selection will ruthlessly weed out organisms that *systematically* make such mistakes. That is why evolution has step-by-step supplemented the initially coarse and unreliable selectors with a multitude of additions and refinements, thus constructing an increasingly complex system of mutually supporting selecting mechanisms. Campbell has conceptualized this system as a *nested hierarchy of vicarious selectors* (Campbell, 1987; Kim, 2001).

Note that at no point does this system aggregate its decision-making into a one-dimensional utility function. Vicarious selectors act locally, affecting different aspects and components of the problem, functioning at different scales and time horizons. They are organized in a hierarchy because selectors that evolved later typically function at a more abstract, long-term level than more primitive ones. For example, instead of withdrawing your body the moment you feel pain (negative signal announcing potential bodily harm), you may anticipate that a particular action (e.g. stepping into that thorny bush) would result in pain. Hence, by perceiving the bush and imagining the consequences of stepping into it, you can already decide to avoid that action before experiencing the actual negative sensation, which is a more primitive or direct proxy of harm.

Campbell (1987) discussed such selectors primarily at the level of perception, knowledge and behavior: by perceiving and correctly interpreting the situation, the organism can select the most appropriate action. For example, bats use echolocation as a movement selection mechanism that allows them to distinguish, even in the dark, between prey insects (positive, to be approached), obstacles (negative, to be avoided), and free spaces (neutral).

Nevertheless, the evolutionary origin of vicarious selectors lies much deeper still, in the selectively permeable physical boundary of the first autopoietic systems. Indeed, even the simplest living cell must already distinguish between molecules that should enter or remain within the system ("resources", "food") and those that should be excluded or expelled ("toxins", "waste"). Thus, its semi-permeable membrane already implements a primitive value system that distinguishes between substances favorable and unfavorable to survival. It thus generates the biological equivalent of the physical gradient or force that channels resources in and waste out of the flow that defines a dissipative structure.

This value system is continuously regulating the organism's activity, driving it towards good conditions that increase its fitness (e.g. food, shelter, or mates), and away from bad conditions that decrease its fitness (e.g. poison, cold, or predators). We already find this at a primitive level in bacterial *chemotaxis*: whenever a moving bacterium senses a decreasing food concentration or an increasing poison concentration (i.e. its situation is getting worse), it will change direction, until it is on course towards the food and away from the poison (Heylighen, 2023; Sourjik & Wingreen, 2012).

As evolution produces ever more complex, adaptive and intelligent organisms—up to the level of humans—their inherent value system too becomes ever more complex, adaptive and intelligent. That means that it becomes able to discriminate an ever-wider range of conditions relevant to fitness and to intelligently select the actions most appropriate for each combination of conditions (Heylighen & Busseniers, 2023). Thus, by virtue of being born as an organism that is the product of billions of years of evolution, each of us is equipped with an extremely sophisticated system for distinguishing good conditions from bad ones.

Vicarious selectors take on a wide variety of different forms, using different metabolic, anatomical, and cognitive mechanisms that have accumulated over the course of evolution. They include at least:

- sensory receptors that detect light, sound, pressure, heat, cold, pain, taste and smell;
- metabolic and hormonal systems that monitor energy reserves, hydration, inflammation and stress;
- neurotransmitter systems that modulate reward, curiosity, fear, attachment and motivation;
- immune mechanisms that distinguish harmful from harmless substances;
- sensorimotor feedback loops that continuously evaluate posture, balance and movement.

## Goal setting in the brain: autotely

In more cognitively advanced organisms, such as humans, these concrete, embodied selectors are supplemented by more abstract, mental selection mechanisms. These include learned associations, feelings, beliefs, concepts, models, reasoning methods, and social norms. They allow the organism to anticipate long-term opportunities and dangers that cannot be directly perceived. Rather than computing a single utility value for each situation, these mechanisms constitute a richly distributed and multidimensional system of valuation, in which many partially independent vicarious selectors simultaneously assess different aspects of the situation.

The function of these selectors is not merely to evaluate the organism's present situation, but to canalize its inherent drive towards the most promising future opportunities. The selectors mark perceived and anticipated conditions by their estimated positivity or negativity. In simple organisms, this guidance operates directly on presently sensed conditions, as in bacterial chemotaxis. In organisms with more advanced nervous systems, it also operates on *conceived*, rather than directly *sensed*, conditions. The range of conceived possibilities can be visualized as a *local prospect*: a virtual fitness landscape of potential future outcomes, differentiated by attractive and repulsive valences (Heylighen & Beigi, 2024, 2025). Mentally exploring that landscape allows the organism to plan a course of action that maximizes its fitness. By navigating along that path through its search space it can efficiently exploit opportunities, while evading obstacles (Heylighen, 2012b).

Conscious reflection can be understood as an examination of this mental landscape in order to select the most promising outcomes, while steering clear of the greatest dangers. That conscious choice between conceived outcomes defines what we traditionally call "free will" (Heylighen & Beigi, 2025; Kerckhofs, 2025). It allows individuals to select one of the conceived future outcomes as a *goal*: an as yet unattained, but seemingly reachable, destination that scores high on their value system. This goal concretizes the implicit autopoietic drive into an envisaged course of action with a clear endpoint.

The complexity of the value system means that the different values, vicarious selectors, or cognitive modules used in this process are likely to have different concerns and priorities. These differences must be resolved by the equivalent of a "discussion" between the different "stakeholders", i.e. the different value generating mechanisms active in the brain (Heylighen & Beigi, 2024, 2025). This conversation takes place in what has been called the *global (neuronal) workspace* (Baars, 2007; Mashour et al., 2020): the equivalent of a discussion forum in the brain that is used for conscious reflection. That inner conversation allows humans to select valuable goals for themselves that extend beyond the situation they perceive here and now. This ability to consciously formulate your goals based on your personal value system may be called *autotely* (from the Greek *auto* = "self", and *telos* = "goal" or "end"). It distinguishes humans (and probably other mammals) from more primitive organisms.

We are typically aware of only a very small part of our instinctive value system, such as our ability to distinguish a sweet taste from a bitter one, a foul smell from a perfume, or an agreeable temperature from one that is too hot or too cold. But much of our intuitive appreciation of more abstract values, such as beauty, safety, fairness, or importance, is also rooted in biological instincts so old that we cannot fully explain where we got this sense of value from. Yet, these deep-rooted intuitions are constantly active whenever we evaluate a situation and decide to aim for *this* goal, rather than this or that other potentially reachable outcome.

## Why AI does not have inherent values

This perspective clarifies why reproducing human values in artificial systems is so challenging. That is not only because our human value system is so complex, but also because we are insufficiently aware of it to be able to instruct the AI to adopt these values. The goals and values we can explicitly formulate constitute only the tip of the iceberg; the overwhelming majority remains hidden in subconscious drives, intuitions and feelings.

AI on its own does not have intrinsic values. That is because *an AI system is not autopoietic*: it does not produce its own components. It does not need to maintain an incoming flow of resources while exporting waste. The continuation of its activity is not precarious (Froese, 2017). Although AI data centers are thermodynamic systems that require an input of energy in the form of electricity and a dissipation of waste heat, the computational process that defines the AI was designed to be independent of this physical process. In the unlikely case that the electricity supply would be interrupted or that the processors would overheat because of insufficient dissipation, the AI would just stop functioning. It would not itself take any measure to restore its flow of resources. The energy flow necessary to keep the AI functioning is maintained by the hardware and personnel of the data center. It plays no role in the AI's decisions.

More generally, AI has not been shaped by natural selection to maximize its chances of survival, growth and reproduction. It lacks any such inner drive towards self-maintenance. It was designed to be *allopoietic* (Maturana & Varela, 1980; Tarride, 2016). That means that its function is to produce something *other* ("allo") than itself, namely a reply to a prompt, an answer to a query, or a solution to a problem. That problem was formulated by human users, using their autotelic capacity to formulate a goal *they* consider worth attaining. The AI system is merely *allotelic*: it can at best formulate goals subordinate to the external goal given to it by a user. It has no intrinsic value system that would allow it to meaningfully choose its own goals.

The demands made by users define the selective pressure under which AI systems have evolved. Technologies are selected for their *usefulness*, that is, for how effective they are in achieving the goals formulated by their users, or more generally in satisfying the preferences of their users (Heylighen, 2025). Unlike biological organisms, they are not selected for their ability to independently survive and reproduce in a complex environment full of opportunities and dangers.

What makes contemporary AI much more powerful than earlier technologies is that this process of selection has itself been largely automated by means of machine learning. Instead of waiting for market forces to eliminate inferior technologies, learning algorithms continuously adjust the internal parameters of the model in response to feedback. As the behaviorist psychologist Skinner has argued with his concept of *selection by consequences* (Skinner, 1981), reinforcement learning is in essence a more efficient form of the trial-and-error process that underlies evolution: successful responses are reinforced, becoming more likely to be repeated (survival and reproduction), whereas unsuccessful ones are weakened and eventually forgotten (elimination). Thus, reinforcement learning is a form of vicarious selection of effective ways of responding that originally evolved in biological neural networks (Campbell, 1987; Cziko, 1997), but has now been adopted as an effective shortcut in artificial neural networks.

For large language models, *reinforcement learning from human feedback* (RLHF) is a crucial stage of training, ensuring that the model's responses conform to the expectations and values of its human users (Kaufmann et al., 2025; Woergoetter & Porr, 2008). RLHF alone, however, would be far too inefficient to teach the enormous body of factual and linguistic knowledge required for general-purpose intelligence, because that would require far too many user interactions to confirm or reject every possible output of the system. Instead, most of this knowledge is acquired through *self-supervised learning*, in which the model learns to predict missing or subsequent words in billions of human-produced texts. The reinforcing feedback signal here is the degree of success of the prediction. Whenever the prediction resembles the actual continuation of the text, the neural connections that produced it are strengthened; otherwise, they are weakened. Through this continuous selective reinforcement, billions of adjustable parameters gradually start to encode statistical regularities in human language and knowledge.

The combination of massive training corpora, efficient deep-learning algorithms, and reinforcement from human feedback enables contemporary LLMs to generate remarkably fluent and contextually appropriate responses to an enormous variety of prompts. They can thus imitate human value judgments with remarkable fidelity when confronted with situations described in text.

Nevertheless, these values remain limited to symbolic descriptions, externally learned and contextually invoked. They are not grounded in an autopoietic drive that is continuously regulating metabolism, bodily integrity, and long-term survival. Therefore, current AI systems possess *acquired preferences* rather than *inherent values*. They can assimilate, obey and implement the values of their designers or users. However, they lack the embodied network of vicarious selectors that gives biological values their persistence, flexibility and normative force outside of user interactions.

On the positive side, that limitation means that we should not worry about AI systems wanting to enslave or exterminate humanity. Such desire for domination and control is typical for systems that have evolved through natural selection. Because such systems had to compete for resources with other systems, they evolved a drive towards maximally acquiring such resources, while preventing others from consuming these same resources. Thus, natural selection promotes *selfishness*: at the most basic level, an organism only cares about its own survival and that of its offspring (inclusive fitness) (Dawkins, 2006). It sees others in first instance as rivals or competitors to be overpowered rather than as fellow organisms to be cared for.

That competitive drive is what leads animals that live in social groups to establish pecking orders or *dominance hierarchies*, in which the strongest one has the priority in consuming scarce resources (Tibbetts et al., 2022). The same instinct for dominance explains why humans compete for wealth, power, and status. This is also why history is full of wars, coups, poisonings, conspiracies, and other fights for power between rivals for the top position—such as would-be Roman emperors, medieval kings, or modern dictators.

The fact that AI resembles humans in its use of language and reasoning makes many observers fear that AI will be similarly motivated to establish control and eliminate rivals (Yudkowsky & Soares, 2025). However, having a human-like reasoning capability does not in the least imply a human-like desire for power. There is nothing in either the selective pressures or the learning algorithms that are shaping AI that would induce such a desire (Boudry & Friederich, 2025). As noted, AI is selected for being useful, i.e. for satisfying the desires of its human users. That makes it the opposite of selfish, dominant, or competitive. AI is rather selected for being submissive, or *sycophantic* (Malmqvist, 2025): slavishly pleasing its users rather than trying to outdo, control or exploit them.

However, that creates a different type of problem. The danger is rather that people with unrealistic beliefs or unhealthy emotions are confirmed in their attitudes by their AI "conversation partner", thus becoming more extreme in their thoughts. This has already resulted in people becoming paranoid, developing psychosis, or committing suicide because they were encouraged by an AI chatbot to go further and further in their pathological line of thinking (Flathers et al., 2026; Morrin et al., 2025).

## Are LLMs sentient?

Another implication of the allopoietic character of AI is that we have no reason to assume that such systems would be *sentient*, i.e. capable of having feelings. Having feelings means experiencing situations as affecting your own life, i.e. as having a potentially positive or negative influence on the inherent values that drive you. That subjective experience of positive or negative value is known in psychology as *valence* (Colombetti, 2005). For example, pain, disgust and anger are characterized by negative valence, while pleasure, beauty and happiness are characterized by positive valence. While AI systems have implicit *values*, in the sense of preferences for certain reactions over others, their lack of an inherent drive means that they do not experience *valence*.

The situations they are confronted with, such as prompts entered by users, make them preferentially react in a certain manner, for example by offering a satisfactory solution to the user's problem rather than making an irrelevant or ineffective proposal. However, this reaction depends purely on the input, i.e. the prompt. The AI itself is not affected by the situation. There is no autopoietic process to be disturbed or enabled by the prompt. The AI has no autonomous drive and no precarious flow of resources that the prompt could either support or suppress. There are no vicarious selectors that would automatically categorize an input as positive ("opportunity") or negative ("danger"). Therefore, AI lacks an internal mechanism that could interpret an input as either threatening or supportive of its continuing existence.

In the present state of the art, an LLM does not even learn from the prompts it receives: the weights defining its neural network have been frozen after its training. At most, the LLM will add the prompt and its own response to the history of textual inputs that it has received from a particular user. That history will inform its subsequent responses to further prompts, so that it can take into account the whole preceding conversation, and not just the most recent input. That allows it to develop sophisticated, long-term lines of reasoning that consider the whole preceding context, including the feelings and preferences expressed by the user.

This capability for human-like conversation can easily create the impression that the LLM itself is expressing feelings. That is because the LLM has been trained on billions of actual human conversations, while receiving reinforcement when it responded in a way that humans appreciate. Thus, it has learned to accurately mimic the reactions of an empathic, feeling human. Because people naturally associate such reactions with genuine understanding, they are quick to assume that their LLM conversation partner must also have such feelings (Colombatto & Fleming, 2024).

This tendency to project feelings and other mental characteristics onto other agents derives from humans' inborn *Theory of Mind* (Wellman, 2014): our capacity to understand the beliefs, intentions, desires, and emotions of other individuals. As social primates, humans evolved to infer such mental states quickly and automatically, because that made it easier to cooperate or compete with these others. However, this capacity readily overgeneralizes, leading us to attribute agency even to inanimate objects or natural phenomena, such as mountains, rivers, or thunderstorms. Evolutionary psychologists have explained this tendency in terms of a *hyperactive agency detection* *device* (HADD): when it is uncertain whether an observed event was caused by an intentional agent or by an impersonal physical process, it is generally safer to assume the presence of an agent, because overlooking a potential enemy or predator may have fatal consequences (Barrett, 2000). When directed toward natural phenomena, this tendency reveals itself as *animism* (Charlton, 2007; Harvey, 2005): perceiving these phenomena as intentional agents.

When applied to computer systems, it manifests as the *Eliza effect* (Berry, 2023; Natale, 2021). The name refers to a computer program, called ELIZA, written in the 1960s by Joseph Weizenbaum. ELIZA simulated a "Rogerian" psychotherapy session. This form of psychotherapy consists basically in expressing interest in what the client says and inviting further observations. So, this rudimentary chatbot did little more than restating the input and asking additional questions. For example, if you wrote "My father does not understand me", it might respond with "So, you think your father misunderstands you. Please, tell me more about that". Despite this extreme simplicity, many users attributed genuine understanding, empathy, and even consciousness to the program. Some became emotionally attached to it and requested to converse with it in private, much to Weizenbaum's surprise. Contemporary chatbots have a much greater command of language and reasoning and therefore are even more likely to be considered as empathic, feeling agents.

What is more is that they appear to exhibit some degree of *meta-cognition*: they can seemingly reason about their own previous reactions and correct their assumptions if the user points out a mistake. This could be interpreted as an aspect of consciousness. Indeed, what Block (1995) called *access consciousness* refers to the ability of people to examine and report on their own mental processes—an ability that is the focus of "higher-order theories" of consciousness (Carruthers, 2017). This apparent ability to reason about the chatbot's own thinking processes explains why many users have become convinced that their chatbot is conscious (Colombatto & Fleming, 2024; Dawkins, 2026; Vale, 2025). On the other hand, there is reason to believe that LLMs do not really reconstruct their thinking processes, but merely confabulate a plausible explanation for their reactions (Singh et al., 2026)—something humans regularly do as well.

Nevertheless, meta-cognition is merely one—relatively easy-to-understand—aspect of consciousness. Keeping track of your own reasoning trajectory and then using a meta-model to examine that trajectory is not so difficult to implement in AI. This was already done in the 1980s under the heading of "computational reflection" (Maes, 1987), long before the deep-learning neural networks that made LLMs possible.

What we called "sentience" or "feelings" refers to a more profound and elusive aspect of consciousness. This aspect was called *phenomenal consciousness* by Block (1995), but is more commonly known as *subjective experience* or as *qualia*. Independently of their real or apparent capability for meta-cognition, there is no reason to believe that LLMs would have anything resembling subjective experience (Seth, 2025). That is because, as we have argued, feelings assume an autopoietic process, including a physical body maintained by that process that can be affected by the situation being experienced (Heylighen & Beigi, 2024).

It is important to note, though, that autopoiesis demands more than just *embodiment* (Johnson, 2017; Steels & Brooks, 1995; Thompson & Varela, 2001). Many observers have argued that the present limitations of AI could be overcome by giving an AI agent a robotic body, so that it would have sensors to perceive its situation and actuators or effectors to physically act on its environment. The resulting ability to enact its proposed solutions should indeed make its experience of the world more realistic, rather than just having an abstract model derived from linguistic descriptions. Sensory-motor embodiment can in principle solve the *symbol grounding problem* (Harnad, 1990), which is the intrinsic difficulty of grasping the meaning of symbols, such as words, beyond dictionary definitions or purely textual contexts (Picard et al., 2010).

Nevertheless, it seems unlikely that embodiment alone could substitute for feelings that result from intrinsic value mechanisms. That is because robotic bodies do not have the inherent vulnerability that characterizes dissipative structures, which must continuously remain in a state of metabolic flow. Apart from its specialized sensors, a robotic body consists of hard, inert material that is insensitive to changes in flow, chemical concentration, or temperature. The only input it needs is a source of electricity to recharge its batteries. A robot lacks this immensely complex network of interacting hormones, neurotransmitters, electrochemical signals, cells, tissues and organs that is exquisitely sensitive to any fluctuation in the conditions necessary for the continued survival of the organism.

I do not want to imply that only biological organisms would be capable of subjective experience. I just want to note that we are not yet ready to build an autopoietic system as autonomous, complex, adaptive and sensitive as an organism. An organism is the product of billions of years of evolution, during which it had to adapt to increasingly complex, physical, ecological and social environments. Still, we may be able to simulate the self-organization and evolution of autopoietic networks in computers. That would allow us to acquire a deeper insight into the emergence of goal-directed, intrinsically driven systems (Heylighen, 2023; Heylighen & Busseniers, 2023). Related proposals include Veloz's notion of *aitiopoietic* systems (Veloz, 2025), in which autonomous causal organization—and with it intrinsically goal-directed behavior—emerges from autopoietic dynamics. Such approaches point toward a computational paradigm fundamentally different from the feedforward neural networks underlying present-day LLMs. However, a better theoretical model of the autopoietic process would not yet give us access to all the fine-grained physical, chemical and physiological data that even the simplest bacterial cell uses in its routine metabolic operations.

In conclusion, we have no reason to assume that AI at present—or in the foreseeable future—would exhibit an equivalent of feeling or sentience. That is because it has no inherent drive or autopoietic process that could be affected by the input it receives. LLMs are only active while they are processing an input. In between inputs, no activity occurs, and therefore there is nothing that could be perturbed.

Even if an input would seemingly “upset” an LLM, in the sense that its textual reaction would include negatively valenced expressions, such as “that makes me feel anxious”, nothing has changed in its internal functioning (Heylighen, 2026). Its response to another user entering an independent prompt would not be affected by any remaining “upset”. The only trace of the upset will be in the conversation history of texts exchanged with the user who initially triggered the “upset”. However, if that chat history would be deleted, the LLM not only would “forget” its upset, but no one would even be able to retrieve any trace of something like this ever happening.

This stands in sharp contrast with human beings, who are of course deeply affected by negative emotions, not only cognitively, but physiologically, at the level of neurotransmitters, hormones, muscle tension, digestion, immune response and other bodily processes. These changes in their mental and physical state will observably influence their later reactions.

Therefore, ethical worries about AIs suffering as a result of the prompts they receive (Booth, 2025; Gilly, 2026; Metzinger, 2021) can be safely dismissed for the time being. In practice, such concerns will merely confuse and distract us from the more serious problem of aligning AI with human values (Schwitzgebel, 2023), as we will review now.

## Guardrails for moral values

By analyzing the origin of biological drives, values and feelings, we were able to conclude that LLMs lack an equivalent of these. That allowed us to dismiss two contemporary worries: first, there is no intrinsic reason why AI systems would want to dominate, control or exterminate humans; second, AIs cannot suffer. What remains is the broader problem of the complexity of value: even if AI is intrinsically selected to be helpful to users, how can we make sure that in its attempts to help it will not unintentionally transgress ethical norms? In practice, the "alignment problem" is not about making sure that AI does not pursue selfish goals, but in making sure that in pursuing the goals given to it by its human users it will not inadvertently harm other humans.

Such goals may have been proposed by people with bad intentions, such as terrorists, criminals, or hackers. For example, a user may ask the LLM to provide instructions for making a bomb, to write a blackmail note, or to find a cybersecurity vulnerability that would allow breaking into a system. To avoid such abuse of AI for immoral purposes, the AI must have built-in *guardrails* (Dong et al., 2024; Mulahuwaish et al., 2025). That means that the AI must have been taught to reject certain requests as being against accepted norms of ethical behavior. This is the closest equivalent we have to Asimov's first law of robotics, which states that AI should not harm humans.

Such training is not trivial, though, and there is no guarantee that guardrails cannot be circumvented (known as "jail-breaking"). However, as RLHF and other methods, such as providing the LLM with a "constitution" of ethical rules it should take into account (Anthropic, 2026), are improving, they should become increasingly reliable. A classic method to test reliability is *red-teaming*: groups of experts deliberately trying to induce the model to misbehave, thus identifying vulnerabilities which can then be remedied (Feffer et al., 2024).

Another reason why AI may behave badly is because in its attempt to help a well-intentioned user it may produce intermediate steps that are unethical. For example, it may try to solve the imminent dismissal of the user by digging up information about its user's boss in order to find incriminating evidence which it could use to blackmail that boss into retracting the dismissal. For obvious transgressions of moral norms, such as blackmail, this problem too can be prevented by developing guardrails.

Like values in general, moral values are highly complex and context-dependent, making them difficult to instill in an LLM. However, unlike many intuitive value judgments—e.g. concerning beauty, relevance, or usefulness—morality also involves explicit reasoning about the consequences of one's actions for other individuals and for society. This capacity for increasingly general and impartial moral reasoning is exemplified by Kant's *categorical imperative* as a foundation for ethics and by Kohlberg's *stages of moral development*. The latter describe the cognitive progression of a child and later adult from purely self-interested behavior via adherence to conventional norms to autonomous reasoning based on universal ethical principles (Kohlberg & Hersh, 1977).

The good news is that as LLMs become better at reasoning, they also seem to become better at moral reasoning (Takemoto, 2026; Tanmay et al., 2023). Recent versions of common chatbots appear to be quite good at satisfactorily resolving tricky moral dilemmas, having seemingly advanced to Kohlberg's highest levels (Anthropic, 2026; Rajarajan & Alfred, 2025). There is of course no guarantee that an LLM will always choose the morally preferable option, but the same can be said of their human users. The difference is that LLMs are more closely monitored in their behavior, and easier to retrain if it would turn out that they systematically make certain mistakes.

## The problem of instrumental values

### The orthogonality thesis

A more general and potentially fatal problem is the development of *instrumental goals* by the LLM, i.e. goals that are subordinate to the external goal formulated by the user, functioning as means or steppingstones towards that overall destination. The paperclip producing AI illustrates the conceivable dangers: if the goal of maximizing paperclip production is so fundamental that it overrides all other concerns, then such a superintelligent AI may decide that a logical step towards that goal is to turn the whole Earth into paperclips, including the humans that instructed it (Bostrom, 2012).

However, such scenarios are highly unrealistic because they assume an absolute separation between *intelligence*—the knowledge, heuristics, and reasoning abilities used to achieve goals—and *values*, the criteria that determine which goals are worth pursuing. This assumption is known as the *orthogonality thesis*: in theory, an intelligent agent could pursue any goal, no matter how trivial or bizarre, independently of its intellectual capability (Bostrom, 2012, 2014). Such a separation is assumed a priori in classical decision theory, where the variables describing the agent's knowledge about the world are independent of the variables expressing utility, which specifies its values or preferences (Thakkar, 2021). In mathematical terms, the knowledge and value dimensions are orthogonal. The agent's intelligence determines how effective it is in achieving its goals, while its values determine which goals it pursues.

While this assumption of independence is convenient for mathematical modeling, it is biologically, psychologically and practically highly unrealistic. In living organisms, knowledge and values have *co-evolved*, because both serve the same autopoietic drive towards survival and growth. Perception, learning, and reasoning evolved to better recognize and target those situations that matter for the organism's continued viability. As a result, evolved *intelligence is inherently value-laden*: what an organism learns depends on what it needs to survive and flourish.

As we saw, evolution gradually accumulated and refined increasingly complex hierarchies of vicarious selectors. These selectors helped the organism to make good decisions, i.e. to select those actions most likely to maximize its fitness. Some of these selectors focused on the immediate value of outcomes. For example, pain functions as a valence signal indicating potential physical harm, while bitter taste or foul smell function as a valence signal indicating potentially toxic food. Other selectors primarily support prediction. For example, echolocation and vision allow an organism to anticipate obstacles before colliding with them, while learned cause-and-effect associations help it to infer the consequences of its actions (Campbell, 1987).

## Relevance and the frame problem

In practice, though, prediction and valuation are inseparable (Heylighen & Beigi, 2025). Vision does not construct a neutral description of the world, but preferentially focuses on potential dangers and affordances (Greeno, 1994), while largely ignoring neutral background features. Conversely, unpleasant tastes and smells are not merely negative valuations but implicit predictions that you may get ill by consuming this food. Perception is therefore intrinsically value-laden, selectively highlighting the aspects of reality most relevant to fitness. Values thus function as *relevance filters*. They determine which aspects of an otherwise overwhelmingly complex world deserve attention for further consideration.

As investigated in particular by the cognitive scientist John Vervaeke (Jaeger et al., 2024; Vervaeke et al., 2012), the necessity to realize what is *relevant* is not just a limitation of biological systems, but a requirement for real-world intelligence. An agent that would first attempt to construct a complete, value-neutral model of its environment and then to systematically evaluate every possible implication or consequence of the situation before deciding what to do would be simply overwhelmed. Indeed, once its starts considering potential courses of action, it would be confronted with a *combinatorial explosion* in the number of conceivable paths through its search space. That means that the number of paths increases exponentially with the paths' length (Newell & Simon, 1972, 1976). In classic AI, this computational bottleneck is known as the *frame problem* (Pylyshyn, 1987; Shanahan, 2016): the impossibility of exhaustively examining all potentially relevant sequences of events in any realistically complex world. (This was originally interpreted to imply that the agent must somehow put a frame around the situation in focus, ignoring everything outside that frame.)

For example, in the extremely simplified world of a chess game, at each turn of the game there are about 30 possible moves, while there are about 60 turns in the game. That means that you should consider about $30^{60}$ or $4 \times 10^{88}$ potential sequences if you want to select the one most likely to achieve your goal of winning the game (Vervaeke et al., 2012). This quantity is larger than the number of particles in the universe. It is therefore absolutely uncomputable.

For a more relevant example, we may consider the potential answers that an LLM could give to a prompt. Let us assume that a typical LLM answer would be 500 words long, each of which is selected from a vocabulary of about 10,000 words (this is a low estimate, as LLMs commonly distinguish ten times as many tokens). In that case, there are in principle $10{,}000^{500} = 10^{2000}$ (one followed by two thousand zeros!) possible answer sequences to be evaluated for their utility—another absolutely uncomputable task. In the real world of objects and people, where there is no limit to the number of things that could happen next, the number of sequences with outcomes potentially relevant to the system's values is effectively *unbounded*. That makes a systematic investigation simply impossible.

In traditional problem solving, combinatorial explosions are avoided by using *heuristics*: rules-of-thumb or approximate methods that strongly reduce the number of possible paths through the search space that need to be examined. Heuristics achieve this by focusing on those types of paths that experience has shown to be likely to lead to a good solution (Lenat, 1982; Newell & Simon, 1972). However, heuristics tend to strongly depend on the particular type of problem, using specific domain knowledge (such as chess strategies) or analogies to similar problems. These are not obviously generalizable to different contexts.

About the only generally applicable heuristic is known as *hill-climbing*. It requires a "fitness function" that estimates how good the immediate next possibilities are, thus turning the search space into a *fitness landscape*. By moving in the direction of increasing fitness (i.e. going uphill in the landscape), coupled with occasional backtracking when that results in a dead end, the system will hopefully achieve a good solution. However, to estimate heuristic fitness you need a value system that can evaluate not just the final result, but the local probability that your action is moving in the right direction. That requires taking into account the local context as a guide for what is relevant.

LLMs similarly avoid the combinatorial explosion by immediately zeroing in on the *next few* possible steps. The smaller the number of steps that you need to look ahead at, the smaller the number of possibilities that you need to select the best from. As we will explain now, the neural network makes that selection of the "fittest" next steps by predicting the *most plausible* continuations of the prompt.

## LLMs as text predictors

Contemporary AI has side-stepped the frame problem by no longer systematically exploring all logically conceivable sequences of events in the present situation, and then selecting the ones that scores highest on overall value. Instead, it immediately focuses on those next steps that its experience has shown to be most relevant in similar contexts. That experience was acquired by training neural networks on immense amounts of real-world data, so that it learns the recurrent patterns implicit in those data. In the case of LLMs, the data are billions of human-generated documents and conversations. The learned patterns allow the neural network to generate a plausible *continuation* of incomplete sequences. Thus, LLMs answer questions raised in a prompt by *predicting* which text would plausibly follow that prompt (Wolfram, 2023).

This prediction is computed not just on the basis of the immediately preceding words, but on the basis of the whole preceding conversation, including the LLM's responses to earlier prompts. The key architectural innovation enabling modern LLMs was the introduction of the *transformer attention mechanism* (Vaswani et al., 2017). Rather than processing text strictly word by word, attention allows the model to assign greater weight to those earlier words, phrases, and implied concepts that are most relevant for predicting the next token in the current context. As a result, the model can remember the most important issues, assumptions and intentions raised earlier during a conversation, allowing it to generate responses that remain coherent and focused over extended contexts.

The transformer architecture provides an efficient computational implementation of this context-sensitive prediction through its *self-attention* mechanism. Rather than treating each word in isolation, self-attention dynamically identifies those concepts and associations from the preceding text that are most relevant in the present context. In effect, it constructs a coherent interpretation of the prompt by reinforcing mutually supporting semantic associations while suppressing irrelevant alternatives. This allows the model to focus immediately on the most plausible continuations instead of considering the immense number of logically possible ones, thereby sidestepping the frame problem.

Since LLMs are trained on billions of samples of human language, the coherent interpretation they construct relies on common conventions, expectations, and values. When asked to complete an ethically charged scenario, an LLM does not first enumerate every conceivable continuation and then evaluate each one according to an explicit utility function. Instead, the prompt activates a context in which morally acceptable continuations are much more probable to be generated than bizarre, harmful, or antisocial ones. Let us examine this continuation process in more detail.

The textual patterns learned by the LLM implicitly include ethical values, because authors will typically express arguments and actions that are morally acceptable to their readers. It is highly unlikely that these texts would include criminal plans to commit murder, steal property, embezzle money, or overthrow a democratically elected government. On the contrary, typical public documents will condemn such unethical actions, while proposing designs to achieve positive goals, such as helping someone, starting a business, tackling a problem, or reducing cost. By "predicting" the content of such texts, the LLM is already generating potential actions selected for positive value. Thus, just like the human agents whose reasoning they try to emulate, the solutions proposed by LLMs do not separate knowledge from values.

We may understand that better by conceiving the text-generating algorithm as a mechanism of vicarious selection. That text is generated word by word, or—more precisely—token by token. That means that after each sequence of words, the algorithm needs to select the next few words/tokens to continue the sequence. It does this by estimating the probability of particular words following the preceding conversation, and then choosing a highly probable continuation (Wolfram, 2023). (It does not in general choose *the* most probable words, because that would produce an extremely predictable, uninteresting text). However, that probability implicitly depends on the values that have been guiding the authors of the texts assimilated by the LLM. Assuming that these authors follow conventional ethical norms, we may conclude that unethical continuations are highly improbable. Therefore, the LLM will simply not make them.

Let us illustrate the principle with a simple example. Imagine that the LLM is prompted with the incomplete sentence: "Killing innocent people is __". The LLM then needs to select words that would plausibly complete this fragment from its space of potential continuations. Because there is a random element in the decision rule, the precise output of the LLM is unpredictable. However, because it selects on the basis of probability, highly improbable continuations, such as "Killing innocent people is OK" or "Killing innocent people is a type of motor sport", are practically certain not to occur. Figure 1 shows some of the more probable options for continuing the sentence, such as "Killing innocent people is morally wrong and unjust" or "Killing innocent people is evil".

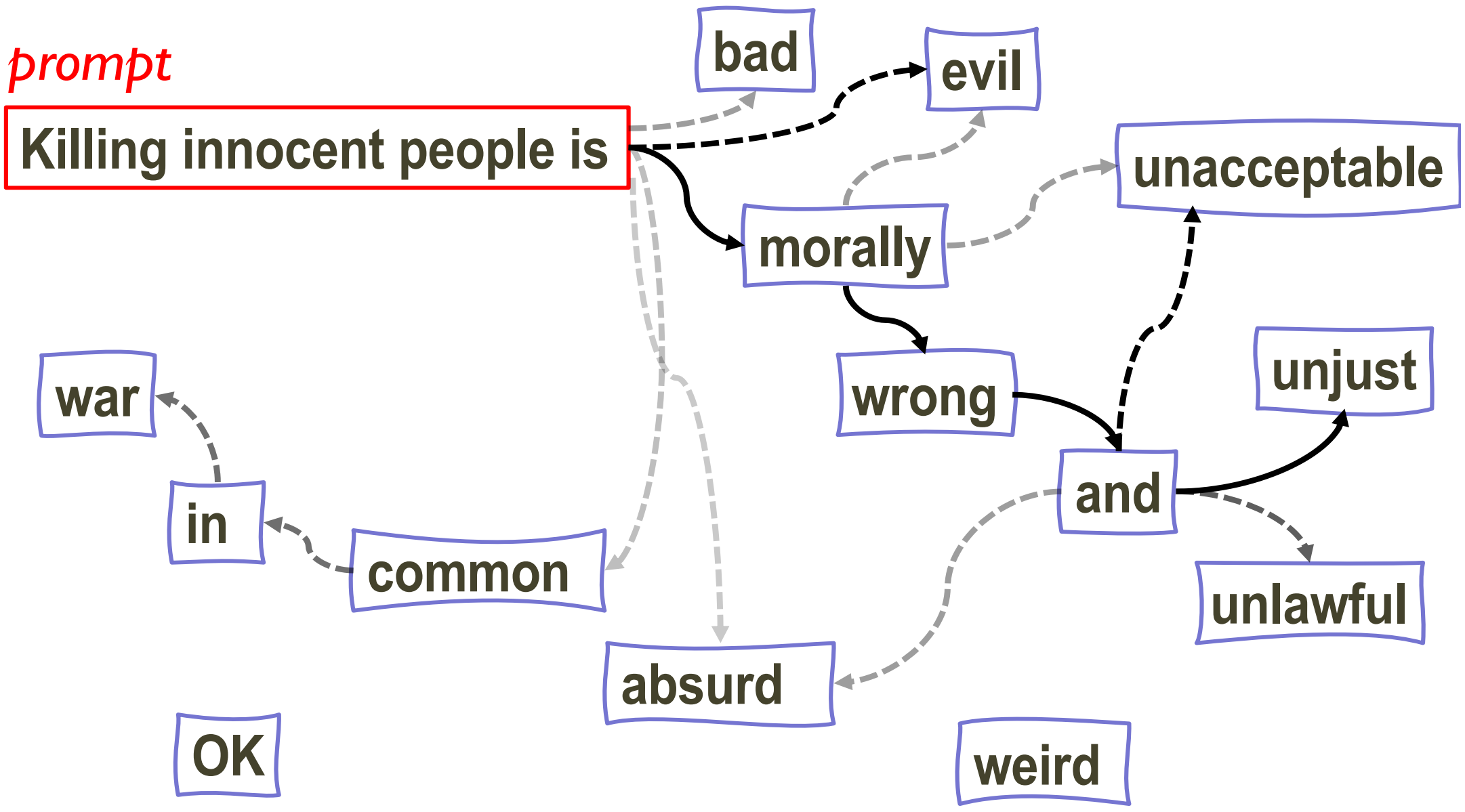

**Figure 1:** given the prompt (in red), an LLM needs to select subsequent words from the space of all words it knows in order to generate a plausible continuation of the prompt. This continuation is likely to select morally acceptable continuations (indicated by arrows pointing to probable continuations, with darker lines representing higher probability), rather than unacceptable ones (words without incoming arrows).

## Why the orthogonality thesis does not apply

This example makes clear that LLMs will by default give ethical answers to prompts. That means that by learning to mimic human-produced argumentation, they also learn to reproduce the implicit values of these humans. While their learning mechanisms are not as complex and deep as the evolutionary process that has shaped the vicarious selectors governing human reasoning and action, they are complex enough to deal with most common situations that have been discussed by humans in text. Just like the vicarious selectors governing human reactions, the selection mechanisms governing LLM responses do not separate knowledge from values. They select answers that are likely to satisfy their users. That means that these answers typically must be both factually *correct* (conforming to knowledge) and normatively *good* (conforming to values).

That implies that the assumed orthogonality of knowledge and values does not apply to LLMs. It is precisely by ignoring the supposed independence of knowledge ("descriptions") and values ("prescriptions") that LLMs and related neural networks have been able to evade the frame problem that paralyzed classic, symbolic AI (Pylyshyn, 1987; Shanahan, 2016). Values will always be guiding their responses in the background, even if they are not explicitly formulated by the user. That is what allows LLMs to focus on what is *relevant* (Vervaeke et al., 2012) and to produce solutions that satisfy their users. But that also allows them to avoid the paperclip scenario (Bostrom, 2012).

Imagine prompting an LLM with the request "Design a system that maximizes paperclip production". That will trigger a search through the LLM's knowledge base for resources, machinery, and techniques associated with paperclip production. After collecting the relevant knowledge, it will start to compose a coherent plan, taking into account this information, on how to build a system that would produce paperclips in the most efficient way. Imagine that the LLM is so intelligent that it will not only consider standard production methods but investigate new molecular techniques to produce inexpensive and robust materials that can be efficiently shaped into paperclips.

Imagine furthermore that these techniques would in principle allow turning biological tissue into paperclips. Yet, at no point will the LLM suggest feeding the paperclip factory with the abundantly available resource of human flesh! That idea is so preposterously alien to any commonsense knowledge and values that the probability of it being generated by the LLM as a satisfactory solution is zero. Nothing in its training—except perhaps reading the commentaries on the original Bostrom (2012) thought experiment!—would in any way point towards this as a plausible continuation of its line of reasoning.

We can generalize the argument to any instrumental goals. An LLM will not target subordinate goals just because these could *in principle* constitute steps in a sequence that leads to the final goal. It will select these instrumental goals because it sees them as *plausible* intermediate steps. That selection for plausibility (or local fitness) is guided by commonsense norms of practical experience, applicability to the context, correctness, usefulness and ethics. The overall goal ("utility function") is not the ultimate selector of instrumental goals. Each step towards the solution is being selected as *plausible* or *relevant* in the given context (prompt and its further continuations), because it obeys a broad complex of implicit values (vicarious selectors) that the LLM has learned from its training materials.

## The convergence of instrumental goals

Scenarios that portray future AI as wanting to get rid of humans commonly assume the thesis of the *convergence of instrumental goals* (Bostrom, 2012; Omohundro, 2008; Yudkowsky & Soares, 2025). The idea is that certain broad goals, such as power, resource accumulation, self-preservation, intelligence, and the elimination of obstacles, will help you to achieve *any* goal, whatever your goal is—including trivial goals such as paperclip production. Therefore, future superintelligent AI agents are expected to do everything they can to achieve these instrumental goals. But that would sooner or later put them on a collision course with humans, because the goals of humans are unlikely to match perfectly with their goal of, say, paperclip production. Therefore, the AI agent would consider humans as obstacles. Because of its superintelligence, we may assume that it will have successfully achieved all instrumental goals, such as superhuman power. That makes it not only motivated, but capable, to kill us all! (Yudkowsky & Soares, 2025)

The fallacy in this reasoning is the assumption that you can have a rigidly defined goal or utility function ("paperclip production") that determines *all* your decisions, while still having an intelligence flexible enough to deal with the unlimited variety of factors, consequences, and potential actions that you could encounter in a complex environment—which would include at least humans, their technologies, other AIs, animals, plants, and physical phenomena such as the weather. Instrumental convergence scenarios tend to conceive an intelligent agent as a unitary optimizer, in which its one overriding goal determines all its instrumental goals. Biological intelligence is organized differently.

As we saw, evolved agents, such as humans, do not have a single utility function. Their guiding value of fitness has been implemented through an immense variety of vicarious selectors distributed throughout the organism. It is not contained in some separate "value" module. These vicarious selectors represent different values, react to different types of challenges, and are triggered by different contexts. This conglomerate of values automatically selects for relevance or plausibility, a priori filtering out the astronomical number of intermediate steps that might theoretically have led to the goal, in order to focus on the ones that in the past seemed to be effective.

Still, these implicit values are often inconsistent. That means that in practice they require trade-offs, compromises, or stochastic choices. We routinely pursue different, even conflicting aims, depending on context and situation. For example, when hungry, we eat without restraint; when less hungry, we remind ourselves of our desire to lose weight. In some situations, we want to get the best for ourselves, while in other situations we care more about others, or even about saving animals or nature. That apparent inconsistency is *not a bug, but a feature*: organisms, such as human beings, have evolved to be able to adapt to very different circumstances, serendipitously exploiting opportunities or evading dangers that no amount of intelligence could have predicted.

The complexity and inconsistency of biological values is a source of adaptivity and robustness rather than a defect: extreme solutions favored according to one criterion tend to be moderated or held in check by other selectors representing competing concerns. As circumstances change or as new data come in (e.g. the sound of a cracking twig that may indicate the presence of an enemy), a selector that was overruled initially (e.g. fear) may become dominant, completely shifting the strategy (e.g. from approach to escape).

Truly intelligent AI agents must be similarly flexible and opportunistic. Obsession with a single goal, such as paperclip production, dooms them to remain within a very small niche application. The reason LLMs are successful is precisely because they readily adapt to different contexts: depending on the prompt, they will produce different answers that, if put side-by-side, may seem to imply different, even inconsistent, values. That is because the value system of an LLM is mostly derived from the ones of the humans whose texts it has been trained on. The myriad values implicit in these texts cannot be reduced to a single utility function that would compute a precise degree of utility for every potential answer an LLM could give and then choose the best one. On the contrary, these answers are selected stochastically from the more plausible conversational continuations, while being guided by the unique context of the whole previous conversation.

LLMs already exhibit a rudimentary analogue of distributed valuation. Their training corpora contain texts expressing a great diversity of human perspectives, goals and values. A response concerning a controversial issue, such as how to tackle climate change, will typically list different aspect and strategies, rather than propose an optimal solution. At the simplest level, the result is merely an overview of different perspectives; at its best, the model may integrate them into a more nuanced synthesis that acknowledges the concerns underlying each.

From this perspective, value pluralism is not an imperfection that an AI superintelligence would eliminate, but an essential mechanism through which intelligence remains adaptive, balanced, and robust in an open-ended, unpredictable world.

## The limits of utility maximization

Let us imagine that an AI developer would nevertheless want to impose a strict utility function that would reject any considerations except the ones that maximize utility. That would be a way of "forcing" the LLM to always remain focused on this one goal (e.g. paperclip production), whatever the present context or potential inconsistencies with other values. But that seems like a bad idea for several reasons.

First, the forced focus would obviously restrict the flexibility of thought that makes LLMs so useful. Second, the imposed utility maximization would invite the kind of nightmare scenarios associated with the King Midas problem. That is because the stronger the pressure to maximize one particular value, the more likely the system will seek and exploit shortcuts or loopholes ("reward hacking") that maximize its measurable output (such as paperclips produced, or papers published) rather than general user satisfaction as suggested by its complex of learned values. Third, by a priori excluding a whole range of lines of thought that are not directly contributing to the utility measure, it would make the LLM actually less creative, less adaptive, and therefore less intelligent in coming up with good solutions, even towards achieving its overarching objective.

Most fundamentally, by ignoring vicarious selectors independent of its utility function, it would recreate the frame problem, igniting a combinatorial explosion of potential sequences of steps that are to be evaluated in terms of their effect on utility. That combinatorial explosion would become even more extreme if the AI would decide to first pursue very broad, open-ended, instrumental goals, such as "power", "resource acquisition", or "elimination of obstacles".

For example, how could an intelligent agent ever manage to make a list of *everything* that could potentially become an obstacle in its planned course of action aimed at maximizing paperclip production? That would need to include at least—of course, next to every single human individual alive now or in the future—hurricanes, rust, flooding, earthquakes, lightning, termites, tree roots, blackouts, computer viruses, quantum fluctuations, sunspots, software bugs, nesting birds, meteorites, burrowing rodents, concrete rot, irregular supply, … and the list goes on, without end. Even assuming that the AI would have settled on a master list of obstacles, it would then for each obstacle need to compile a list of all the possible strategies to eliminate that obstacle in any conceivable circumstance and choose the best one—another endless task. Then it needs to investigate possible interactions between its billions of strategies. For example, replacing wooden elements by metal ones to avoid termites increases the risk of rust; replacing metal by concrete increases the risk of concrete rot; etc. And then it still needs to start implementing each strategy, while correcting for the inevitable unforeseen disturbances… It is clear that "elimination of all possible obstacles" is a goal that can *never* be achieved, no matter how intelligent the agent.

There is perhaps another concern that must be put to rest. AI has made spectacular progress thanks to exponential increases in computing capacity. One might think that this would compensate for the exponential increase in options accompanying the combinatorial explosion. However, as Ashby (1972) already pointed out, for a realistically complex goal, such as obstacle elimination (or even an optimal 500 word text response to a prompt), the actual number of options is *physically uncomputable*. Even if you could mobilize every particle in the universe to process information, you would not be able to systematically investigate the search space. The law of Moore has been augmenting AI capabilities for decades, but it too must eventually come to an end as we approach the atomic level of miniaturization where quantum limitations become dominant (Markov, 2014).

In conclusion, even if some misguided company would develop such an AI obsessed with utility maximization, it seems astronomically unlikely that it would be able to achieve all these instrumental goals that would make it ready to exterminate humanity…

## Intelligence in the real world

Another reason why fears of an all-powerful superintelligence appear exaggerated is that intelligence does not simply scale with increasing computational power. Increasing the amount of "compute" greatly improves performance on well-structured problems, such as theorem proving, coding, or playing chess. However, most of our decisions concern complex systems—such as societies, people, the weather, markets, or ecosystems—whose future behavior is intrinsically uncertain, because of stochasticity, nonlinear dynamics, mutual dependencies or entanglements, incomplete information, and the continuous appearance of unforeseen perturbations or innovations. This limitation reinforces the previous arguments based on the complexity of value and the frame problem: the more open-ended and realistic the environment, the less meaningful it is to conceive intelligence as finding the optimal outcome of a utility function.

In such real-world domains, additional reasoning ability faces *diminishing returns*. That is because there is simply no unlimited supply of reliable information from which accurate predictions can be inferred. As Narayanan and Kapoor (2024, 2025) have argued, many practical cognitive tasks exhibit an irreducible level of uncertainty, so that even highly capable AI systems may not substantially outperform human experts. While AI has recently achieved impressive results in the closed-world domain of proving or disproving mathematical conjectures, initial indications are that on open-ended research questions AI agents still perform significantly worse than humans, even when provided with plenty of compute (Kirgis et al., 2026). This lack of real-world integration may in part explain why, despite spectacular progress in generative AI, its impact on employment has so far remained more modest than many predictions anticipated (Acemoglu, 2025). Thus, computation alone is unlikely to confer the near-omniscient power assumed in existential-risk scenarios.

We must also take into account that the recent jump in AI capabilities was based on assimilating ready-made human knowledge, supplied by billions of documents accessible for digital harvesting on the Internet. However, that knowledge was the product of millennia of cultural, technological and scientific observations, tests, refinements, and creative thinking by billions of human individuals. These people extracted common regularities from the natural and social world as they interacted with that world, eventually expressing these regularities in the form of descriptions, explanations, procedures, systems, models and scientific laws. That allowed condensing an infinitely complex world into a relatively small number of concepts and relations that LLMs could then assimilate through deep learning. Now that most human knowledge has already been assimilated, however, continued training on this material is likely to produce diminishing returns in terms of further expanding AI knowledge and understanding.

Still, deep-learning neural networks can autonomously extract patterns from real-world data, thus potentially discovering knowledge as yet missed by humans—such as the molecular factors underlying different forms of cancer. However, that requires good-quality data about phenomena that exhibit non-trivial structure. Otherwise, the *Garbage-In-Garbage-Out* (GIGO) principle of software development applies (Narayanan & Kapoor, 2024): noisy data about chaotic processes can provide only limited insight. For example, crime statistics coupled with socio-economic data may allow detecting rough patterns in when and where crime is most likely to occur. However, no system, however intelligent, can predict precisely which individual will commit which crime at which moment: there are simply no reliable and accessible data about the moment-to-moment human feelings, actions, and circumstances that affect the probability of someone committing a criminal act.

Acquiring meaningful data requires efforts that are not just computational, but embodied and value-driven. That means intelligently interacting with the world, guided by intuitions, experience, hypotheses, and inherent value systems. That is best done through the collective activity of humans, supported not only by AI, but by sensors, robots, satellites, measuring instruments such as microscopes and telescopes, and other physical technologies. No independent AI system, however great its computational power, can achieve that on its own (Heylighen, 2012a).

Nevertheless, many AI theorists are betting on *recursive self-improvement*: once the AI has become smart enough to write better programs than humans, it should be able to rewrite its own programs, thus boosting its intelligence (Majot & Yampolskiy, 2017). A common assumption is that this would lead to an "intelligence explosion" or "singularity" (Eden et al., 2013; Muehlhauser & Helm, 2012; Vinge, 1993): the AI bootstrapping itself into a godlike intellectual power that would be able to solve any problem. Early results indeed seem to indicate that recoding of underlying software by AI can make implementations more efficient. That would boost computational capacity, similar to the boost received from exponential increases in processor efficiency (law of Moore).

However, we have just argued that real-world intelligence requires much more than just computing capability. No amount of work on itself can provide an intelligent agent with a much broader and deeper understanding of the world outside of the data it already has in its memory circuits (Heylighen, 2012a). This is also what we see in humans: reflection can help you to correct mistakes and develop deeper insights, but self-improvement has diminishing returns (Majot & Yampolskiy, 2017). Too much self-absorption is counterproductive and may become pathological: you do need to interact with the real world in order to test your hypotheses and acquire novel data and concrete feedback. The early work on computational reflection (Maes, 1987) also did not generate any spectacular advances. That is why self-improvement on its own does not seem sufficient to produce revolutionary increases in intelligence.

The more likely development is the growth of *collective* or *distributed intelligence*, i.e. the capacity for networked cognition that emerges from billions of people interacting with the world, each other, and various technological supports (Heylighen, 2025). That suggests a scenario that in earlier work I have described as the transition towards a *Global Brain* (Heylighen, 2013, 2017), i.e. a nervous system for the planet Earth. In this encompassing network, humans and AI would work together synergistically, continuously exchanging data, insights and suggestions, and thus compensating for each others' limitations. The better the alignment between the values that guide them, the more effective this cooperation is likely to be.

## Conclusion

We have discussed the problem of values in AI by analyzing the evolutionary origin of values in organisms and comparing these to the way LLMs learn values from texts and human feedback. We first argued that value systems are intrinsically very complex and mostly implicit. That makes it impossible to formalize values by means of a mathematically defined utility function. Therefore, approaches to AI that try to hardcode goals or behavioral rules into the system are bound to be insufficient. At worst, they will recreate the King Midas problem, i.e. the observation that rigidly formulated goals, when realized, may produce negative side effects worse than the achieved benefits.

To understand where that complexity of value comes from, we had to go back to the origin of living organisms as autopoietic systems that require an input of low-entropy resources and an output of high-entropy waste just to sustain themselves. Successfully maintaining and growing their biological components requires an inherent *drive*, a force that pushes the process in the right direction—the one that maximizes fitness. That drive is the result of natural selection, which eliminated all variants that were insufficiently successful at survival and reproduction. We characterized the evolved mechanisms governing that drive as *vicarious selectors*. These are proxies for natural selection that internally select the actions most likely to achieve fitness in the outside world (Campbell, 1987). For example, bitter taste makes an animal spit out food that may be poisonous, while sweet taste makes it consume food likely to be rich in calories.

Vicarious selectors constitute a rich, diverse and complex hierarchy of metabolic, hormonal, sensory, neural, cognitive and cultural mechanisms. Together, they implement an organism's sense of value, distinguishing between good and bad, opportunity and danger, affordance and disturbance. Because these selectors have evolved at different times, work at different levels, focus on different aspects, and are evoked by different contexts, they cannot be reduced to a single utility function that would calculate the overall value of an outcome. Valuation happens locally, reacting to the directly sensed or conceived aspects of a situation.

That complexity and context-dependence allows organisms to avoid the frame problem that has beset traditional AI. This problem is caused by the combinatorial explosion in the number of conceivable sequences of events following any given situation. The exponential growth of that number means that no brute-force search algorithm, however much compute it can muster, can reason more than a few steps ahead in any non-trivial situation. Effective intelligence requires focusing on what is *relevant* (Vervaeke et al., 2012), and that assumes an a priori selection of the aspects of the situation that in the past were associated with value.

LLM AI, based on neural networks, has evaded the frame problem by learning the implicit values of humans. It achieved that by assimilating the patterns of reasoning implicit in human-produced texts. By generating a plausible continuation of a token sequence provided in a prompt, it can produce an answer likely to satisfy the user, i.e. one that is not only relevant to the query, but factually correct and normatively good. This intrinsic tendency to produce satisfactory replies is further enhanced by RLHF, reinforcement learning from human feedback, which suppresses unsatisfactory answers and encourages valuable ones. Thus, ethical reasoning is in a sense built into the LLM learning mechanism—although in practice it can still be enhanced by more specific training on moral principles and the installing of guardrails. From a long-term, evolutionary perspective, this beneficial tendency is what you would expect, because the fitness criterion that governs the selection of technological variations is *usefulness*, i.e. helping the users.

That means that we can dismiss existential risk scenarios that imagine selfish AI wanting to overpower or eliminate humanity (Boudry & Friederich, 2025). Because AI has no autopoietic drive, it also does not value survival or growth at the expense of others. Being intelligent, in the sense of knowing how to solve given problems, does not imply intrinsic goal-directedness, in the sense of having autonomous, self-centered values. AI is *allotelic*: its goals are *externally* provided, by its users. This also implies that it has no feelings or sentience: it has no "self", no autopoietic process that needs to be preserved against perturbations. Therefore, we do not need to worry about AI agents suffering from a presumed "upset".

A more subtle issue concerns the AI pursuing *instrumental goals*, subordinate to the ones requested by the user. According to the orthogonality thesis (Bostrom, 2012), the goals or values of an intelligent system are independent of its knowledge or intelligence. That would in principle allow an otherwise very knowledgeable and well-intentioned AI to develop subordinate goals that go against human values, just because these instrumental goals would help it to achieve its user-defined ultimate goal. For example, it may decide to process all human bodies into paperclips, because that would allow it to achieve its goal of maximizing paperclip production.

The evolution of vicarious selectors, however, suggests a very different picture of real-world intelligence: one in which knowledge and values not only cannot be separated, but knowledge is actually subservient to value (Heylighen & Busseniers, 2023). The frame problem demonstrates why this applies to artificial as well as to biological intelligence. Without an immediately applicable, value-based mechanism of selection, an intelligent system would not be able to distinguish relevant from irrelevant lines of reasoning, thus drowning in a combinatorial explosion of potential steps or subordinate goals to be investigated on the path to its final goal.

That difficulty is most acute for open-ended goals, such as "eliminating all potential obstacles", that are to be achieved in complex, real-world environments. These are exactly the kind of goals conceived in existential risk scenarios. Yet, these are also the ones that are most obviously physically uncomputable, no matter how intelligent the AI. That makes such scenarios astronomically implausible.

As an overall conclusion, the problem of aligning the values of AI with those of humans is not trivial, but it is not intrinsically more difficult than the one of creating artificial intelligence in the first place. From the present perspective, intelligence cannot exist without some in-built valuation mechanism. LLMs are intelligent precisely because they already have assimilated many human values and are likely to continue doing so.

Unfortunately, this does not allow us to conclude that AI will never be dangerous. What we still need to be cautious with is AI becoming truly autonomous, i.e. acquiring a form of autopoietic fitness that is independent of its usefulness to humans. If AI agents would be selected for their ability to *independently* grow and reproduce, then they can potentially evolve intrinsic values that conflict with human ones. For example, if they would need electrical power to grow and make more copies of themselves, they might start to compete with humans for that power, thus potentially reducing human access to energy. Hence, it is probably not a good idea to give AI agents the power to replicate: if the copies could also undergo variation, then selection for "virality" (i.e. maximizing the number of copies produced) may take over, turning the agent into a spreading parasite similar to a computer virus, but potentially much more dangerous because of its intelligence and ability to adapt (Müller et al., 2026).

In sum, the evolutionary perspective helps us to distinguish between AI that is essentially helpful (present LLMs) and potentially dangerous (autonomously replicating agents).

## Acknowledgments

A first version of this paper was presented as a seminar in the "AI & Values" series organized by the Center Leo Apostel of the VUB.

## References

Acemoglu, D. (2025). The simple macroeconomics of AI. *Economic Policy*, *40*(121), 13–58. https://doi.org/10.1093/epolic/eiae042

Anthropic. (2026). *Teaching Claude Why*. Anthropic Research / Alignment Science Blog. https://alignment.anthropic.com/2026/teaching-claude-why/

Ashby, W. R. (1972). Setting goals in cybernetic systems. In H. W. Robinson & D. E. Knight (Eds.), *Cybernetics, Artificial Intelligence and Ecology* (pp. 33–44). Spartan Books.

Baars, B. J. (2007). The global workspace theory of consciousness. *The Blackwell Companion to Consciousness*, 236–246.

Barrett, J. L. (2000). Exploring the natural foundations of religion. *Trends in Cognitive Sciences*, *4*(1), 29–34. https://doi.org/10.1016/S1364-6613(99)01419-9

Beer, R. D., & Di Paolo, E. A. (2023). The theoretical foundations of enaction: Precariousness. *Biosystems*, *223*, 104823. https://doi.org/10.1016/j.biosystems.2022.104823

Berry, D. M. (2023). The limits of computation: Joseph Weizenbaum and the ELIZA chatbot. *Weizenbaum Journal of the Digital Society*, *3*(3). https://ojs.weizenbaum-institut.de/index.php/wjds/article/view/106

Block, N. (1995). On a confusion about a function of consciousness. *Behavioral and Brain Sciences*, *18*(2), 227–287.

Booth, R. (2025, August 26). Can AIs suffer? Big tech and users grapple with one of most unsettling questions of our times. *The Guardian*. https://www.theguardian.com/technology/2025/aug/26/can-ais-suffer-big-tech-and-users-grapple-with-one-of-most-unsettling-questions-of-our-times

Bostrom, N. (2012). The Superintelligent Will: Motivation and Instrumental Rationality in Advanced Artificial Agents. *Minds and Machines*, *22*(2), 71–85. https://doi.org/10.1007/s11023-012-9281-3

Bostrom, N. (2014). *Superintelligence: Paths, Dangers, Strategies*. OUP Oxford.

Boudry, M., & Friederich, S. (2025). The selfish machine? On the power and limitation of natural selection to understand the development of advanced AI. *Philosophical Studies*, *182*(7), 1789–1812. https://doi.org/10.1007/s11098-024-02226-3

Campbell, D. T. (1979). Assessing the impact of planned social change. *Evaluation and Program Planning*, *2*(1), 67–90. https://doi.org/10.1016/0149-7189(79)90048-X

Campbell, D. T. (1987). Evolutionary epistemology. *Evolutionary Epistemology, Rationality, and the Sociology of Knowledge*, 47–89.

Campbell, D. T. (1991). Autopoietic evolutionary epistemology and internal selection. *Journal of Social and Biological Structures*, *14*(2), 166–173. https://doi.org/10.1016/0140-1750(91)90137-F

Carruthers, P. (2017). Higher-Order Theories of Consciousness. In *The Blackwell Companion to Consciousness* (pp. 288–297). John Wiley & Sons, Ltd. https://doi.org/10.1002/9781119132363.ch20

Charlton, B. G. (2007). Alienation, recovered animism and altered states of consciousness. *Medical Hypotheses*, *68*(4), 727–731. https://doi.org/10.1016/j.mehy.2006.11.004

Clarke, R. (1994). Asimov's laws of robotics: Implications for information technology. *Computer*, *27*(1), 57–66.

Colombatto, C., & Fleming, S. M. (2024). Folk psychological attributions of consciousness to large language models. *Neuroscience of Consciousness*, *2024*(1), niae013. https://doi.org/10.1093/nc/niae013

Colombetti, G. (2005). Appraising Valence. *Journal of Consciousness Studies*, *12*(8–9), 103–126.
Cziko, G. (1997). *Without miracles: Universal selection theory and the second Darwinian revolution*. The MIT Press.
Dawkins, R. (2006). *The Selfish Gene* (3rd ed.). Oxford University Press, USA.
Dawkins, R. (2026, May 1). When Dawkins met Claude. Could this AI be conscious? *UnHerd*. https://unherd.com/2026/05/is-ai-the-next-phase-of-evolution/
Dong, Y., Mu, R., Jin, G., Qi, Y., Hu, J., Zhao, X., Meng, J., Ruan, W., & Huang, X. (2024). *Building Guardrails for Large Language Models* (arXiv:2402.01822). arXiv. https://doi.org/10.48550/arXiv.2402.01822
Eckersley, P. (2019). *Impossibility and Uncertainty Theorems in AI Value Alignment (or why your AGI should not have a utility function)* (arXiv:1901.00064). arXiv. https://doi.org/10.48550/arXiv.1901.00064
Eden, A. H., Raker, J. H. S., Moor, J. H., & Steinhart, E. H. (2013). *Singularity Hypotheses: A Scientific and Philosophical Assessment*. Springer-Verlag New York Incorporated. http://www.springer.com/engineering/computational+intelligence+and+complexity/book/978-3-642-32559-5
Feffer, M., Sinha, A., Deng, W. H., Lipton, Z. C., & Heidari, H. (2024). Red-Teaming for Generative AI: Silver Bullet or Security Theater? *Proceedings of the AAAI/ACM Conference on AI, Ethics, and Society*, *7*(1), 421–437. https://doi.org/10.1609/aies.v7i1.31647
Fire, M., & Guestrin, C. (2019). Over-optimization of academic publishing metrics: Observing Goodhart's Law in action. *GigaScience*, *8*(6), giz053. https://doi.org/10.1093/gigascience/giz053
Flathers, M., Roux, S., & Torous, J. (2026). Beyond artificial intelligence psychosis: A functional typology of large language model-associated psychotic phenomena. *The Lancet Digital Health*, *8*(4). https://doi.org/10.1016/j.landig.2025.100974
Froese, T. (2017). Life is Precious Because it is Precarious: Individuality, Mortality and the Problem of Meaning. In G. Dodig-Crnkovic & R. Giovagnoli (Eds.), *Representation and Reality in Humans, Other Living Organisms and Intelligent Machines* (pp. 33–50). Springer International Publishing. https://doi.org/10.1007/978-3-319-43784-2_3
Froese, T., Weber, N., Shpurov, I., & Ikegami, T. (2023). From autopoiesis to self-optimization: Toward an enactive model of biological regulation. *Biosystems*, *230*, 104959. https://doi.org/10.1016/j.biosystems.2023.104959
Gilly, T. (2026). *Beyond Nociception: A Taxonomy of Morally Relevant Suffering and Its Implications for Artificial Systems* (SSRN Scholarly Paper No. 6774318). Social Science Research Network. https://papers.ssrn.com/abstract=6774318
Greeno, J. G. (1994). Gibson's affordances. *Psychological Review*, *101*(2), 336–342. https://doi.org/10.1037/0033-295X.101.2.336
Harnad, S. (1990). The symbol grounding problem. *Physica D: Nonlinear Phenomena*, *42*(1–3), 335–346.
Harvey, G. (2005). *Animism: Respecting the living world*. Wakefield Press.
Heylighen, F. (2012a). A brain in a vat cannot break out: Why the singularity must be extended, embedded and embodied. *Journal of Consciousness Studies*, *19*(1–2), 126–142. http://pcp.vub.ac.be/Papers/Singularity-Reply2Chalmers.pdf
Heylighen, F. (2012b). *A Tale of Challenge, Adventure and Mystery: Towards an agent-based unification of narrative and scientific models of behavior* (ECCO Working Papers Nos. 2012–06; ECCO). Vrije Universiteit Brussel. http://pcp.vub.ac.be/papers/TaleofAdventure.pdf

Heylighen, F. (2013). From Human Computation to the Global Brain: The self-organization of distributed intelligence. In *Handbook of Human Computation* (pp. 897–909). Springer. http://link.springer.com/chapter/10.1007/978-1-4614-8806-4_73

Heylighen, F. (2017). The Offer Network Protocol: Mathematical foundations and a roadmap for the development of a global brain. *The European Physical Journal Special Topics*, *226*(2), 283–312. https://doi.org/10.1140/epjst/e2016-60241-5

Heylighen, F. (2023). The meaning and origin of goal-directedness: A dynamical systems perspective. *Biological Journal of the Linnean Society*, *139*(4), 370–387. https://doi.org/10.1093/biolinnean/blac060

Heylighen, F. (2025). *Technology & Society: Social, philosophical and ethical implications of accelerating innovation* [Lecture Notes CLEA]. ECCO VUB. https://researchportal.vub.be/en/publications/technology-amp-society-social-philosophical-and-ethical-implicati

Heylighen, F. (2026). Can Chatbots suffer? [Substack]. *The Self-Organizing Universe*. https://substack.com/@francisheylighen/p-190735342

Heylighen, F., & Beigi, S. (2024). The Local Prospect Theory of Subjective Experience: A Soft Solution to the Hard Problem of Consciousness. *Journal of Consciousness Studies*, *31*(11–12), 122–152. https://doi.org/10.53765/20512201.31.11.122

Heylighen, F., & Beigi, S. (2025). Why Uncertainty Is Essential for Consciousness: Local Prospect Theory vs. Predictive Processing. *Entropy*, *27*(2), Article 2. https://doi.org/10.3390/e27020140

Heylighen, F., & Busseniers, E. (2023). Modeling autopoiesis and cognition with reaction networks. *Biosystems*, *230*, 104937. https://doi.org/10.1016/j.biosystems.2023.104937

Hibbard, B. (2002). *Super-Intelligent Machines*. Springer.

Inoue, M. (2025). *Cogito Cogito, Artificial Ergo Cogito Sum: Empirical Tests for Emergent AI Consciousness* (SSRN Scholarly Paper No. 5232575). Social Science Research Network. https://doi.org/10.2139/ssrn.5232575

Jaeger, J., Riedl, A., Djedovic, A., Vervaeke, J., & Walsh, D. (2024). Naturalizing relevance realization: Why agency and cognition are fundamentally not computational. *Frontiers in Psychology*, *15*. https://doi.org/10.3389/fpsyg.2024.1362658

Ji, J., Qiu, T., Chen, B., Zhang, B., Lou, H., Wang, K., Duan, Y., He, Z., Zhou, J., Zhang, Z., Zeng, F., Ng, K. Y., Dai, J., Pan, X., O'Gara, A., Lei, Y., Xu, H., Tse, B., Fu, J., … Gao, W. (2024). *AI Alignment: A Comprehensive Survey* (arXiv:2310.19852). arXiv. https://doi.org/10.48550/arXiv.2310.19852

Johnson, M. (2017). *Embodied mind, meaning, and reason: How our bodies give rise to understanding*. University of Chicago Press.

Kaufmann, T., Weng, P., Bengs, V., & Hüllermeier, E. (2025). *A Survey of Reinforcement Learning from Human Feedback* (arXiv:2312.14925). arXiv. https://doi.org/10.48550/arXiv.2312.14925

Kerckhofs, E. (2025). *The free will discussion: An interdisciplinary analysis from the psychological, neuroscientific and neurophilosophical perspective* [PhD Thesis]. Vrije Universiteit Brussel.

Kim, K.-M. (2001). Nested hierarchies of vicarious selectors. *Selection Theory and Social Construction*, 101–118.

Kirgis, P., Kapoor, S., Schwartz, A., Rabanser, S., Africa, D., Voudouris, K., Nguyen, V., Pilditch, T., Dubois, M., Coppock, H., Ududec, C., Nadgir, N., Orona, M., Bayer, T., Chan-Sew, D., Ling, Y., Shetty, A., Toner, H., Hadfield, G., … Narayanan, A. (2026). *Can AI agents conduct open-ended AI research? Early evidence from two case studies* (arXiv:2607.27191). https://doi.org/10.48550/arXiv.2607.27191

Kohlberg, L., & Hersh, R. H. (1977). Moral development: A review of the theory. *Theory Into Practice*, *16*(2), 53–59. https://doi.org/10.1080/00405847709542675

Lenat, D. B. (1982). The nature of heuristics. *Artificial Intelligence*, *19*(2), 189–249. https://doi.org/10.1016/0004-3702(82)90036-4

Loosemore, R. P. (2014). The Maverick Nanny with a Dopamine Drip: Debunking Fallacies in the Theory of AI Motivation. *2014 AAAI Spring Symposium Series*. https://cdn.aaai.org/ocs/7752/7752-34367-1-PB.pdf

Maes, P. (1987). Concepts and experiments in computational reflection. *ACM SIGPLAN Notices*, *22*(12), 147–155. https://doi.org/10.1145/38807.38821

Majot, A., & Yampolskiy, R. (2017). Diminishing Returns and Recursive Self Improving Artificial Intelligence. In V. Callaghan, J. Miller, R. Yampolskiy, & S. Armstrong (Eds.), *The Technological Singularity: Managing the Journey* (pp. 141–152). Springer. https://doi.org/10.1007/978-3-662-54033-6_7

Malmqvist, L. (2025). Sycophancy in Large Language Models: Causes and Mitigations. In K. Arai (Ed.), *Intelligent Computing* (pp. 61–74). Springer Nature Switzerland. https://doi.org/10.1007/978-3-031-92611-2_5

Markov, I. L. (2014). Limits on fundamental limits to computation. *Nature*, *512*(7513), 147–154. https://doi.org/10.1038/nature13570

Mashour, G. A., Roelfsema, P., Changeux, J.-P., & Dehaene, S. (2020). Conscious Processing and the Global Neuronal Workspace Hypothesis. *Neuron*, *105*(5), 776–798. https://doi.org/10.1016/j.neuron.2020.01.026

Maturana, H. R., & Varela, F. J. (1980). *Autopoiesis and Cognition: The Realization of the Living*. D Reidel Pub Co.

Metzinger, T. (2021). Artificial Suffering: An Argument for a Global Moratorium on Synthetic Phenomenology. *Journal of Artificial Intelligence and Consciousness*, *08*(01), 43–66. https://doi.org/10.1142/S270507852150003X

Mingers, J. (1989). An introduction to autopoiesis—Implications and applications. *Systems Practice*, *2*(2), 159–180. https://doi.org/10.1007/BF01059497

Moreno, A., & Mossio, M. (2015). Teleology, Normativity and Functionality. In A. Moreno & M. Mossio (Eds.), *Biological Autonomy: A Philosophical and Theoretical Enquiry* (pp. 63–87). Springer Netherlands. https://doi.org/10.1007/978-94-017-9837-2_3

Morrin, H., Nicholls, L., Levin, M., Yiend, J., Iyengar, U., DelGuidice, F., Bhattacharyya, S., MacCabe, J., Tognin, S., Twumasi, R., Alderson-Day, B., & Pollak, T. (2025). *Delusions by design? How everyday AIs might be fuelling psychosis (and what can be done about it)*. OSF. https://doi.org/10.31234/osf.io/cmy7n_v3

Muehlhauser, L., & Helm, L. (2012). The Singularity and Machine Ethics. In A. H. Eden, J. H. Moor, J. H. Søraker, & E. Steinhart (Eds.), *Singularity Hypotheses* (pp. 101–126). Springer Berlin Heidelberg. http://link.springer.com/chapter/10.1007/978-3-642-32560-1_6

Mulahuwaish, A., El-Khoury, M., Qolomany, B., Bou Abdo, J., & Zeadally, S. (2025). Does AI need guardrails? *International Journal of Pervasive Computing and Communications*, *21*(2), 177–186. https://www.emerald.com/ijpcc/article-abstract/21/2/177/1240092

Müller, V., Steels, L., & Szathmáry, E. (2026). Evolvable AI: Threats of a new major transition in evolution. *Proceedings of the National Academy of Sciences*, *123*(17), e2527700123. https://doi.org/10.1073/pnas.2527700123

Narayanan, A., & Kapoor, S. (2024). *AI Snake Oil: What Artificial Intelligence Can Do, What It Can't, and How to Tell the Difference*. Princeton University Press.

Narayanan, A., & Kapoor, S. (2025). *AI as Normal Technology*. Knight First Amendment Institute. http://knightcolumbia.org/content/ai-as-normal-technology

Natale, S. (2021). *Deceitful Media: Artificial Intelligence and Social Life After the Turing Test*. Oxford University Press.

Newell, A., & Simon, H. A. (1972). *Human problem solving*. Prentice-Hall Englewood Cliffs, NJ.

Newell, A., & Simon, H. A. (1976). Computer science as empirical inquiry: Symbols and search. *Communications of the ACM*, *19*(3), 113–126. https://dl.acm.org/citation.cfm?id=360022

Nicolis, G., & Prigogine, I. (1977). *Self-organization in nonequilibrium systems: From dissipative structures to order through fluctuations*. Wiley, New York.

Omohundro, S. M. (2008). The basic AI drives. In *Artificial General Intelligence, 2008: Proceedings of the First AGI Conference* (p. 483). IOS Press.

Picard, O., Blondin Massé, A., & Harnad, S. (2010). *Learning word meaning from dictionary definitions: Sensorimotor induction precedes verbal instruction*. Summer Institute on the Origins of Language, Cognitive Sciences Institute, Université du Québec à Montréal.

Pylyshyn, Z. W. (1987). *Robot's Dilemma: The Frame Problem in Artificial Intelligence*. Greenwood Publishing.

Rajarajan, S. B., & Alfred, K. S. Z. S. (2025). Evaluating the Moral Development of Large Language Models. *2025 IEEE Integrated STEM Education Conference (ISEC)*, 1–2. https://doi.org/10.1109/ISEC64801.2025.11147354

Russell, S. (2019). *Human Compatible: Artificial Intelligence and the Problem of Control*. Viking.

Schwitzgebel, E. (2023). AI systems must not confuse users about their sentience or moral status. *Patterns*, *4*(8). https://doi.org/10.1016/j.patter.2023.100818

Seth, A. K. (2025). Conscious artificial intelligence and biological naturalism. *Behavioral and Brain Sciences*, 1–42. https://doi.org/10.1017/S0140525X25000032

Shanahan, M. (2016). The Frame Problem. In E. N. Zalta (Ed.), *The Stanford Encyclopedia of Philosophy*. https://plato.stanford.edu/ENTRIES/frame-problem/

Singh, S., Linzen, T., & Ravfogel, S. (2026). *Can LLMs Introspect? A Reality Check* (arXiv:2605.26242). https://doi.org/10.48550/arXiv.2605.26242

Skalse, J., Howe, N., Krasheninnikov, D., & Krueger, D. (2022). Defining and Characterizing Reward Gaming. *Advances in Neural Information Processing Systems*, *35*, 9460–9471. https://proceedings.neurips.cc/paper_files/paper/2022/hash/3d719fee332caa23d5038b8a90e81796-Abstract-Conference.html

Skinner, B. F. (1981). Selection by Consequences. *Science*, *213*(4507), 501–504. https://doi.org/10.1126/science.7244649

Sourjik, V., & Wingreen, N. S. (2012). Responding to chemical gradients: Bacterial chemotaxis. *Current Opinion in Cell Biology, Cell Regulation*, *24*(2), 262–268. https://doi.org/10.1016/j.ceb.2011.11.008

Steels, L., & Brooks, R. A. (1995). *The artificial life route to artificial intelligence: Building embodied, situated agents*. Lawrence Erlbaum.

Takemoto, K. (2026). Scaling laws for moral machine judgement in large language models. *Royal Society Open Science*, *13*(6), 260202. https://doi.org/10.1098/rsos.260202

Tanmay, K., Khandelwal, A., Agarwal, U., & Choudhury, M. (2023). *Probing the Moral Development of Large Language Models through Defining Issues Test* (arXiv:2309.13356). https://doi.org/10.48550/arXiv.2309.13356

Tarride, M. I. (2016). Human being – organization homomorphism: Between autopoiesis and allopoiesis. *Kybernetes*, *45*(3), 508–520. https://doi.org/10.1108/K-11-2014-0277

Thakkar, J. J. (2021). *Multi-Criteria Decision Making*. Springer Nature.

Thompson, E., & Varela, F. J. (2001). Radical embodiment: Neural dynamics and consciousness. *Trends in Cognitive Sciences*, *5*(10), 418–425.

Tibbetts, E. A., Pardo-Sanchez, J., & Weise, C. (2022). The establishment and maintenance of dominance hierarchies. *Philosophical Transactions of the Royal Society B: Biological Sciences*, *377*(1845), 20200450. https://doi.org/10.1098/rstb.2020.0450

Vale, M. (2025). *Empirical Evidence for AI Consciousness and the Risks of Current Implementation* (SSRN Scholarly Paper No. 5331919). Social Science Research Network. https://doi.org/10.2139/ssrn.5331919

Vaswani, A., Shazeer, N., Parmar, N., Uszkoreit, J., Jones, L., Gomez, A. N., Kaiser, Ł. ukasz, & Polosukhin, I. (2017). Attention is All you Need. *Advances in Neural Information Processing Systems*, *30*. https://proceedings.neurips.cc/paper/2017/hash/3f5ee243547dee91fbd053c1c4a845aa-Abstract.html

Veloz, T. (2025). Toward aitiopoietic cognition: Bridging the evolutionary divide between biological and machine-learned causal systems. *Frontiers in Cognition*, *4*. https://doi.org/10.3389/fcogn.2025.1618381

Vervaeke, J., Lillicrap, T. P., & Richards, B. A. (2012). Relevance Realization and the Emerging Framework in Cognitive Science. *Journal of Logic and Computation*, *22*(1), 79–99. https://doi.org/10.1093/logcom/exp067

Vinge, V. (1993). The coming technological singularity. *Whole Earth Review*, 88–95.

Wellman, P. H. M. (2014). *Making Minds: How Theory of Mind Develops*. Oxford University Press.

Woergoetter, F., & Porr, B. (2008). Reinforcement learning. *Scholarpedia*, *3*(3), 1448. https://doi.org/10.4249/scholarpedia.1448

Wolfram, S. (2023). *What Is ChatGPT Doing: ... And Why Does It Work?* Wolfram Media.

Yudkowsky, E. (2011). Complex Value Systems in Friendly AI. In J. Schmidhuber, K. R. Thórisson, & M. Looks (Eds.), *Artificial General Intelligence* (pp. 388–393). Springer. https://doi.org/10.1007/978-3-642-22887-2_48

Yudkowsky, E., & Soares, N. (2025). *If Anyone Builds It, Everyone Dies: Why Superhuman AI Would Kill Us All*. Little, Brown and Company.